\documentclass[11pt,a4paper]{article}
\usepackage[utf8]{inputenc}
\usepackage{amsmath}
\usepackage{placeins}
\usepackage{url}
\usepackage{natbib}
\usepackage{color,xcolor,setspace,geometry}
\setcitestyle{authoryear,open={(},close={)}}
\usepackage{cite}
\usepackage{amsfonts}
\usepackage[normalem]{ulem}
\usepackage{pifont}
\usepackage{colortbl}
\usepackage{hyperref}

\usepackage{arydshln}
\usepackage{multirow}
\usepackage{caption}
\usepackage{subcaption}

\usepackage{tikz}
\usetikzlibrary{shapes}
\usetikzlibrary{plotmarks}
\usetikzlibrary{tikzmark,matrix,calc}
\usetikzlibrary{arrows.meta,calc,decorations.markings,math,arrows.meta}

\usepackage{amssymb}
\usepackage{multirow}
\usepackage{graphicx}
\usepackage{rotating}
\usepackage{setspace}
\usepackage{threeparttable}
\usepackage{authblk}

\usepackage{marvosym}
\usepackage{bbm}
\usepackage{tikz}
\usetikzlibrary{arrows.meta}
\usepackage{graphicx}
\usepackage{soul}

\usepackage{appendix}
\usepackage{bm}
\usepackage{graphicx}   
\usepackage{booktabs}   

\begin{document}
\begin{spacing}{1.5}	

\title{Discounted Sales of Expiring Perishables: Challenges for Demand Forecasting in Grocery Retail Practice}

\author[$\diamond$ $\star$]{David Winkelmann}
\author[$\S$]{Theresa Elbracht}
\author[$\ddag$]{Jonas Brenker}
\author[$ $]{Arnold Gerzen}

\affil[$\diamond$]{\small Department of Empirical Methods\\ Bielefeld University, Bielefeld, Germany}
\affil[$\S$]{Department of Management Science \& Business Analytics\\ Bielefeld University, Bielefeld, Germany}
\affil[$\ddag$]{Department of Management Information Systems\\  Paderborn University, Paderborn, Germany}
\affil[$ $]{}

\affil[$\star$]{Corresponding author: david.winkelmann@uni-bielefeld.de}

\maketitle
\noindent
\normalsize

\begin{abstract}
Grocery retailers frequently apply price discounts to stimulate demand for expiring perishables. However, integrating these discounted sales into future demand forecasts presents a significant challenge. This study investigates the effectiveness of incorporating a fixed share of these sales as \textit{regular} demand into the forecast, as commonly applied in practice. We employ a two-step regression approach on data from a major European grocery retailer, covering over 1,700 products across 676 stores. We reveal that forecasts underestimate actual demand for most SKUs when discounted sales occur. This residual uplift effect is significantly influenced by the number of sales at reduced prices. Our findings underscore the necessity for more precise approaches to integrate discounted sales into demand forecasts, thereby preventing excess inventory and the associated economic and environmental impacts of spoilage in the grocery sector.

%
\end{abstract}

\newpage
\section{Introduction}
\label{sec:introduction}
Grocery retailers offer a variety of products (stock keeping units, SKUs), including perishables with limited shelf lives, such as dairy and chilled fresh goods. Operating in a competitive environment with narrow profit margins, retailers face high customer expectations regarding product availability \citep{kuijpers2018reviving}. Consequently, they address high service levels \citep{ulrich2021distributional}, which require maintaining substantial inventories. This, however, increases the risk of spoilage, leading not only to financial losses but also to environmental impacts. The inherent complexity of managing inventories \citep{winkelmann2022dynamic} is underscored by roughly 11 million tons of food wasted annually in Germany, including the retail sector \citep{bmleh2025national}.

In response to this issue, major retailers, along with the Federal Ministry of Agriculture, Food and Regional Identity, have pledged to halve food waste by 2030 \citep{bmleh2025national}. A common strategy to achieve this goal involves applying price reductions to items nearing expiration, typically indicated by discount stickers, to minimise unsold inventory and recover part of the product value. As consumer willingness to pay decreases with reduced shelf life, such discounts indeed have the potential of a promotional effect \citep{tsiros2005effect}.

A key challenge in inventory management is accurately forecasting customer demand to make informed replenishment decisions. While demand forecasting is a well-established topic in both academic literature and practice \citep{fildes2022retail}, many retailers still rely on basic methods, such as point estimates and time series models, often enhanced by qualitative assessments from decision-makers \citep{ge2019retail}. Although advanced machine learning methods are available, they typically require extensive data inputs, frequently unavailable in practice \citep{fildes2022retail}. Most demand forecasting approaches use features such as product price, marketing campaigns, and, in particular, historical demand data \citep{ulrich2021distributional}.

\begin{figure}[htp!]
    \centering
    
   \resizebox{0.5\textwidth}{!}{
\begin{tikzpicture}
    \tikzset{
        double arrow/.style={
            -{Implies[length=4mm, width=6mm]},
            double distance=2.5pt,
            thick,
            line cap=round,
        },
        text label/.style={
            font=\small,
            align=center
        }
    }
    
    \draw[] (1,6) rectangle (-1.1,7.1);
    \node[text label, font=\Large] at (0,6.8) {Excess};
    \node[text label, font=\Large] at (0,6.3) {Inventory};

    \draw[double arrow, bend left=40] (1.5,6.5) to (4,4.5);

    \draw[] (3,3) rectangle (5.1,4.1);
    \node[text label, font=\Large] at (4.05,3.8) {Price};
    \node[text label, font=\Large] at (4.05,3.3) {Discount};
    
    \draw[double arrow, bend left=40] (4,2.5) to (1.5,0.5);

    \draw[] (1,1) rectangle (-1.1,-0.1);
    \node[text label, font=\Large] at (-0.05,0.7) {Enhanced};
    \node[text label, font=\Large] at (-0.05,0.2) {Sales};

    \draw[double arrow, bend left=40] (-1.5,0.5) to (-4,2.5);

    \draw[] (-3,3) rectangle (-5.1,4.1);
    \node[text label, font=\Large] at (-4.05,3.8) {Biased};
    \node[text label, font=\Large] at (-4.05,3.3) {Forecast};
    
    \draw[double arrow, bend left=40] (-4,4.5) to (-1.5,6.5);
    \end{tikzpicture}
    }
    \caption{Illustration of the problem description.}
    \label{fig:cycle}
\end{figure}
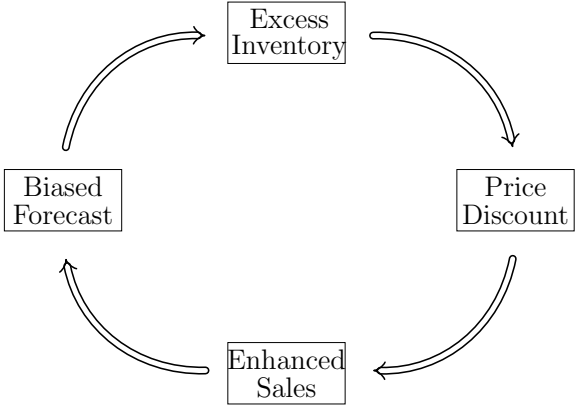

In the context of sales at a reduced price (discounted sales), retailers face the challenge of appropriately incorporating these sales into demand forecasts for future periods. Since some customers may purchase solely due to the discount, fully integrating these increased sales figures can lead to overestimating customer demand, resulting in excess inventory. This surplus necessitates further price reductions, creating a reinforcing cycle illustrated in Figure~\ref{fig:cycle}. Breaking this cycle requires accurately integrating discounted sales into demand forecasts. In particular, retailers must distinguish between purchases regardless of the discount (\textit{regular demand}) and those driven by the reduced price only. As individual purchase motivations are unknown, retailers often assume a portion of sales is independent of discounts and include this share in future demand forecasts.

This study examines the business case of a major European grocery retailer that uses a simplified rule, assuming a fixed share of discounted sales is independent of the discount, and includes only this portion as a feature in demand forecasts. However, the actual share likely varies across products and stores. We explore this issue by analysing the impact of discounted sales on forecast accuracy across different SKUs. Using data from our business partner on inventory levels, forecasts, and (discounted) sales for SKUs in the category \textit{refrigerated fresh products} across multiple stores, we employ a two-step regression approach to quantify the promotional effect of discounts. Our findings confirm the impact of discounted sales on forecasting accuracy with variation across SKUs, highlighting the need for a more precise approach. This analysis forms the basis for improving demand forecasts by more accurately incorporating discounted sales of expiring perishables.

\section{Data}
\label{sec:data}

Our analysis utilises a dataset provided by our business partner. It encompasses 1,705 SKUs across 676 stores over 92 business days from August to October 2024. Each entry corresponds to a single SKU in a specific store on a given day, detailing the stock level at the beginning of each day, the forecasted and actual sales, and the number of sales made at a reduced price. While we lack detailed product-level descriptions, each SKU is hierarchically categorised into one of six supercategories (e.g.\ dairy, cheese, meat), and further divided into up to seven subcategories per supercategory (e.g.\ yoghurt, fresh milk, and cream under dairy). We reduce the dataset to those 1,132 SKUs with at least 100 entries and 50 periods with sales at a discounted price. This results in approximately 20 million entries, with around 3\% involving discounted sales.

Table~\ref{tab:data} provides a sample extract from the dataset with three example rows. Both store and SKU identifiers have been anonymised by our business partner and are represented by indices. In the first row, SKU 10 in store 1 had 12 items in stock on September 22. On that day, two items were sold, both at a discounted price, even though the forecasted demand was only 0.521. A similar example is provided in the last row, where actual sales again exceeded the forecast, but both discounted and regular sales occurred. Conversely, the second row depicts a situation where only one (regular) sale was made out of five available items, closely matching the forecast of 0.736, with no discounts applied.

\begin{table}[htp!]
    \caption{Extract of data covering store/SKU ID, date, and sales-related variables}
    \label{tab:data}
    \centering
    \scalebox{1}{
    \begin{tabular}{r@{\hspace{5pt}}r@{\hspace{8pt}}c@{\hspace{8pt}}lc@{\hspace{5pt}}c@{\hspace{5pt}}c@{\hspace{5pt}}c}
    \hline\noalign{\smallskip}
        Store & SKU & Date & Weekday & Stock & Forecast & Sales & Discounted Sales \\
        \noalign{\smallskip}\hline\noalign{\smallskip}
        1 & 10 & 2024-09-22 & Friday & 12 & 0.521 & 2 & 2 \\
        34 & 579 & 2024-09-25 & Wednesday & 5 & 0.736 & 1 & 0 \\
        676 & 842 & 2024-10-22 & Tuesday & 3 & 0.343 & 3 & 2 \\
        \noalign{\smallskip}\hline\noalign{\smallskip}
    \end{tabular}}
\end{table}

\section{Model}
\label{sec:model}
To assess the uplift effect of sales at a reduced price on customer demand, i.e.\ the additional demand that can be attributed to the presence of discounts, we implement a two-step regression approach: First, we estimate baseline customer demand during periods without discounted sales. Second, we analyse the dependence between discounted sales and the residual lift, i.e.\ the difference between forecasted and actual sales, in periods involving sales at a reduced price. To obtain an initial indication of whether this issue warrants further investigation, we choose a basic linear regression approach. More advanced methods, such as structural equation models \citep{ullman2012structural}, could potentially provide additional insights if further explanatory variables were available in the dataset.

\textbf{Step 1: Estimating Baseline Demand.} To distinguish the effect of discounted sales from the existing forecast used by our business partner, we divide the observation period for each SKU $i \in I$ into two disjunct subsets: days without any discounted sales $t \in T$ and remaining days involving at least one discounted sale $\Tilde{t} \in \Tilde{T}$. We then set up linear regression models for each SKU to explain $\text{\textit{Sales}}_t$, denoting the number of items sold on day $t$. As covariates, we include the daily forecast provided by our business partner ($\text{\textit{Forecast}}_{t}$), the available stock at the beginning of the day ($\text{\textit{Stock}}_t$), and dummy variables for weekdays $\mathbbm{I}\{\text{\textit{Weekday}}_t = w\}$ for Monday ($w = 1$) to Sunday ($w = 7$). The model is formulated as follows:
\begin{equation}
    \text{E}(\text{\textit{Sales}}_t) = \sum\limits_{w=1}^7 \beta_w \cdot \mathbbm{I}\{\text{\textit{Weekday}}_t = w\} + \beta_8 \cdot \text{\textit{Forecast}}_t + \beta_9 \cdot \text{\textit{Stock}}_t.
    \label{eq:reg1}
\end{equation}
By estimating the parameters of the regression model in Equation~\ref{eq:reg1} using ordinary least squares (OLS), we predict sales $\widehat{\text{\textit{Sales}}}_{\Tilde{t}}$ for days $\Tilde{t}$ without considering the uplift effect from discounted sales. The residuals $\delta$ are then calculated as:
\begin{equation}
    \delta_{\Tilde{t}} = \text{\textit{Sales}}_{\Tilde{t}} - \widehat{\text{\textit{Sales}}}_{\Tilde{t}}.
    \label{eq:res}
\end{equation}

\textbf{Step 2: Estimating Promotional Uplift.}
In the second step, we evaluate the impact of sales at a reduced price on forecast accuracy measured by the residual uplift. A positive residual ($\delta_{\Tilde{t}} > 0$) indicates that actual sales during periods involving discounted sales were underestimated, suggesting the presence of an uplift effect. We again set up linear regression models for each SKU using the same explanatory variables as in Equation~\ref{eq:reg1}, now extended by the number of discounted sales $\text{\textit{DS}}_{\Tilde{t}}$, to explain the residuals calculated in Equation~\ref{eq:res} for each day $\Tilde{t}$:
\begin{equation}
    \delta_{\Tilde{t}} =  \sum\limits_{w=1}^7 \gamma_w \cdot \mathbbm{I}\{\text{\textit{Weekday}}_{\Tilde{t}} = w\} + \gamma_8 \cdot \text{\textit{Forecast}}_{\Tilde{t}} + \gamma_9 \cdot \text{\textit{Stock}}_{\Tilde{t}} + \gamma_{10} \cdot \text{\textit{DS}}_{\Tilde{t}} + \epsilon_{\Tilde{t}}.
    \label{eq:reg2}
\end{equation}
We are primarily interested in the estimated parameter value $\hat{\gamma}_{10}$, quantifying the marginal effect of an additional discounted sale on residuals $\delta$. Statistically significant positive effects of discounted sales would suggest that each additional sale at a discounted price increases the residual uplift. Further, we analyse the variation of this parameter across different SKUs to determine the validity of the assumption that the same share of discounted sales occurs regardless of the discounted price for all SKUs.

\section{Results}
\label{sec:results}
We evaluate 1,132 SKUs using the methodology outlined in Section~\ref{sec:model}, which involves estimating the baseline regression model (Equation~\ref{eq:reg1}), calculating residuals (Equation~\ref{eq:res}), and assessing the impact of discounted sales (Equation~\ref{eq:reg2}). Results are obtained for 1,130 SKUs; the estimation failed for the remaining two as not all parameters required for the prediction could be estimated in periods without discounted sales due to missing values for some weekdays. We start by fitting the baseline regression model for each SKU and calculating residuals. We denote the average residual for SKU $i$ by $\Bar{\delta}_i$. To mitigate the influence of outliers caused by data inaccuracies, our analysis focuses on the central 95\% interval of estimates for $\delta$.

\begin{figure}[htp!]
    \centering
    \begin{subfigure}[b]{0.49\linewidth}
        \includegraphics[width=\linewidth]{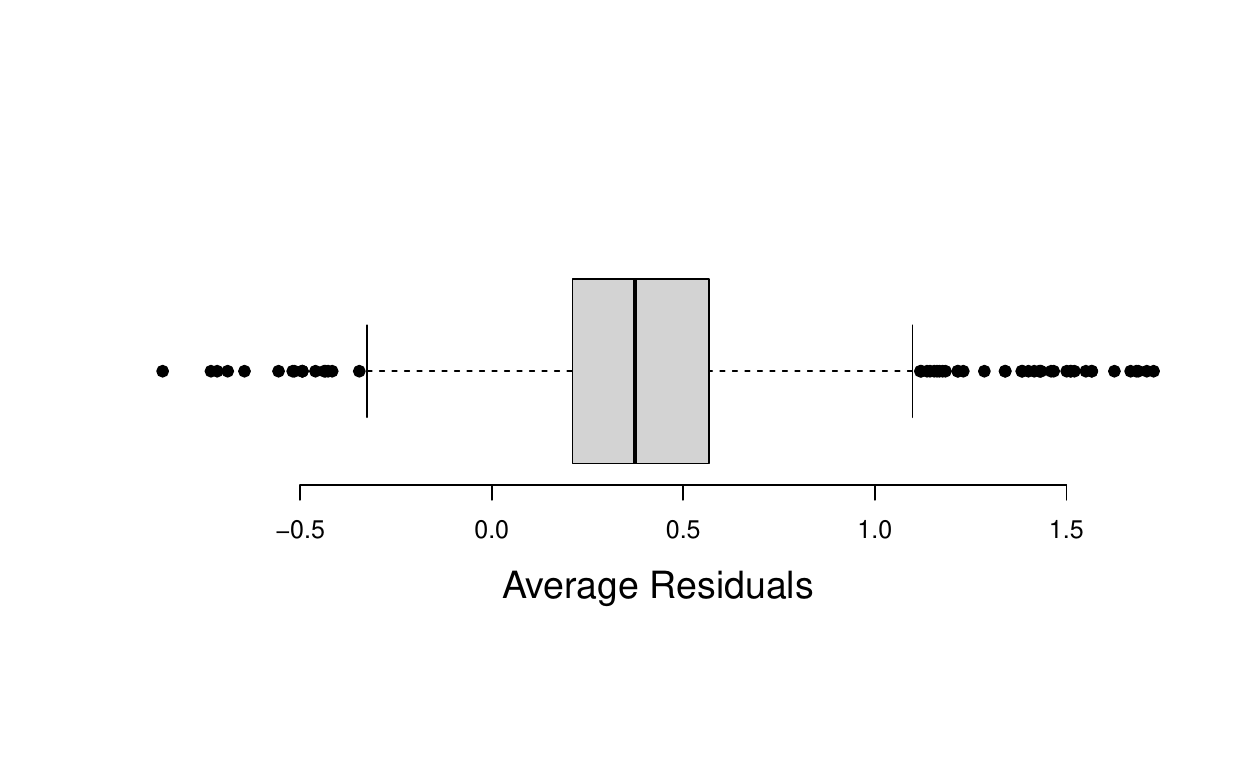}
        \caption{Boxplot of the average residuals between forecasted and actual sales in periods involving discounted sales for different SKUs.}
        \label{fig:residuals}
    \end{subfigure}
    \hfill
    \begin{subfigure}[b]{0.49\linewidth}
        \includegraphics[width=\linewidth]{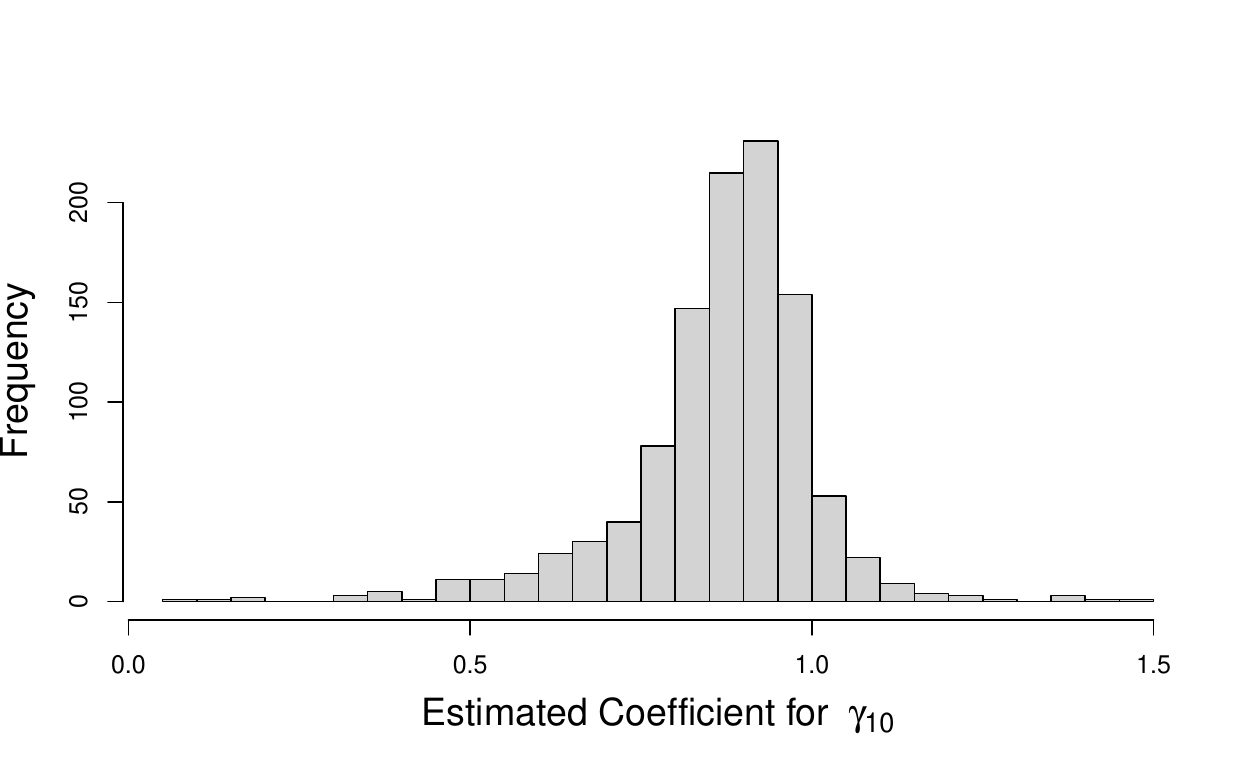}
        \caption{Histogram of estimated coefficients $\hat{\gamma}_{10}$ for various SKUs. For illustrative reasons, 12 values smaller than 0 and larger than 1.5 are excluded.}
        \label{fig:gamma}
    \end{subfigure}
    \caption{Illustration of average residuals and estimated coefficients $\hat{\gamma}_{10}$.}
\end{figure}

Figure~\ref{fig:residuals} shows a boxplot of the average residuals for the remaining 1,077 SKUs. On average, customer demand is underestimated ($\Bar{\delta} > 0$) in periods with discounted sales for approximately 92.2\% of the SKUs; the mean value of average residuals is 0.41. This supports the hypothesis that sales increase when price discounts are applied. We then examine the effect of the number of discounted sales on the residuals. Although customers typically tend to avoid items with a very short remaining shelf life, which could result in a negative estimated coefficient $\hat{\gamma}_{10}$ despite the discounted price, we do not observe a statistically significant negative effect for any SKU. In contrast, a positive effect is observed for 1,067 SKUs, of which 1,044 are statistically significant, confirming the promotional impact of discounted sales. All effects are evaluated at the 5\% significance level using t-tests. Figure~\ref{fig:gamma} presents a histogram of the estimated coefficients $\hat{\gamma}_{10}$, indicating a similar magnitude of the effect for many SKUs. However, there is variation among certain SKUs, warranting further investigation.

\section{Conclusion}
\label{sec:conclusion}
Grocery retailers often use discount stickers to stimulate demand for expiring perishables and reduce spoilage in their inventory. However, this complicates demand forecasting: integrating discounted sales into demand forecasts without accurately accounting for the share of sales driven solely by the discount can result in overestimating future demand and, consequently, excess inventory. While retailers typically rely on some rule of thumb, this study offers initial insights into the uplift effect of such discounts. Our baseline regression model underestimates demand in periods involving discounted sales. Additionally, the number of sales at a reduced price significantly increases this effect for most SKUs. Since the effect varies in magnitude across different SKUs, our study emphasises the need for tailored approaches to effectively incorporate discounted sales for various SKUs. 

Our findings highlight the need for further investigation. Applying more advanced machine learning techniques, such as random forests or neural networks, could help overcome the limitations of the linear regression models used in this study and potentially provide more sophisticated results. In particular, models that are able to estimate conditional average treatment effects, such as tree-based methods \citep{athey2016recursive}, are better suited to address the research questions examined in this study. Additionally, incorporating further explanatory variables that are not available in our dataset, for instance, the location of the store, could improve model accuracy by accounting for contextual information such as holidays. Furthermore, access to data on the number of items exhibiting a discount sticker (rather than just the number of discounted sales), as well as information on the availability of substitutes and complements \citep{karakul2008analytical}, could also offer deeper insights into consumers' tendencies to purchase expiring items at discounted prices. Finally, if information on the remaining shelf life of an item were available, the measure of discounted sales could be directly linked to the risk of spoilage, providing detailed information about the effectiveness of discounted sales as a strategic tool to reduce waste.

\newpage
\bibliographystyle{apalike} 
\bibliography{library}

@misc{kuijpers2018reviving,
  title={{Reviving Grocery Retail: Six Imperatives}},
  author={Kuijpers, Dymfke and Simmons, Virginia and Van Wamelen, Jasper},
  journal={McKinsey \& Company},
  year={2018},
  howpublished = {\url{https://www.mckinsey.com/industries/retail/our-insights/reviving-grocery-retail-six-imperatives}}
}

@article{ulrich2021distributional,
  title={Distributional regression for demand forecasting in e-grocery},
  author={Ulrich, Matthias and Jahnke, Hermann and Langrock, Roland and Pesch, Robert and Senge, Robin},
  journal={European Journal of Operational Research},
  volume={294},
  number={3},
  pages={{ }831-842},
  year={2021},
  publisher={Elsevier}
}

@article{tsiros2005effect,
  title={The effect of expiration dates and perceived risk on purchasing behavior in grocery store perishable categories},
  author={Tsiros, Michael and Heilman, Carrie M},
  journal={Journal of Marketing},
  volume={69},
  number={2},
  pages={114-129},
  year={2005},
  publisher={SAGE Publications Sage CA: Los Angeles, CA}
}

@article{fildes2022retail,
  title={Retail forecasting: research and practice},
  author={Fildes, Robert and Ma, Shaohui and Kolassa, Stephan},
  journal={International Journal of Forecasting},
  volume={38},
  number={4},
  pages={1283-1318},
  year={2022},
  publisher={Elsevier}
}

@article{ge2019retail,
  title={Retail supply chain management: a review of theories and practices},
  author={Ge, Deng and Pan, Yi and Shen, Zuo-Jun and Wu, Di and Yuan, Rong and Zhang, Chao},
  journal={Journal of Data, Information and Management},
  volume={1},
  number={1},
  pages={45-64},
  year={2019},
  publisher={Springer}
}

@article{winkelmann2022dynamic,
      title={Dynamic stochastic inventory management in e-grocery retailing}, 
      author={David Winkelmann and Matthias Ulrich and Michael Römer and Roland Langrock and Hermann Jahnke},
      year={2024},
      journal={arXiv preprint arXiv:{}2205.06572},
}

@article{karakul2008analytical,
  title={Analytical and managerial implications of integrating product substitutability in the joint pricing and procurement problem},
  author={Karakul, Mustafa and Chan, Lap Mui Ann},
  journal={European Journal of Operational Research},
  volume={190},
  number={1},
  pages={179--204},
  year={2008},
  publisher={Elsevier}
}

@misc{bmleh2025national,
  author       = {{Federal Ministry of Agriculture, Food and Regional Identidy}},
  title        = {{National Strategy for Food Waste Reduction}},
  year         = {2025},
  howpublished = {\url{https://www.bmleh.de/EN/topics/food-and-nutrition/food-waste/national-strategy-for-food-waste-reduction.html}}
}

@article{ullman2012structural,
  title={Structural Equation Modeling},
  author={Ullman, Jodie B and Bentler, Peter M},
  journal={Handbook of Psychology},
  volume={2},
  year={2012},
  publisher={Wiley Online Library}
}

@article{athey2016recursive,
  title={Recursive partitioning for heterogeneous causal effects},
  author={Athey, Susan and Imbens, Guido},
  journal={Proceedings of the National Academy of Sciences},
  volume={113},
  number={27},
  pages={7353--7360},
  year={2016},
  publisher={National Academy of Sciences}
}

\end{spacing}
\end{document}